# The Culture of Exploratory Experimentation at CERN and the Future Colliders

Grigoris Panoutsopoulos[1]

## The complex relationship between theory and experiment

We have become witnesses, as of late, to the development of an expansive and fertile debate on the prospects of future colliders, like the upcoming Future Circular Collider (FCC) at CERN[2], both within the confines of the scientific community, as well as outside of them. Within this framework, one frequently encounters a simplistic and outdated view of how science evolves, used mostly as an argumentative basis against such projects. By adhering to a theory-dominated view of science, these arguments place theory at center stage, while experimentation and instrumentation are both relegated to supporting roles, used solely for the purposes of testing a conjecture which has been posited by theory. In other words, in this regard, any future large-scale infrastructure is considered as a questionable scientific investment in the absence of solid theoretical predictions that can be tested at the energy levels achievable by the new colliders.

This view has undergone a radical revision within the field of history and philosophy of science, with a multitude of studies showing that the role of the experiment is quite a bit more involved and multi-faceted. As far back as the 80's and 90's, we have seen the development of historiographical currents that contested a theory-dominated view of science, as it had mostly been represented through the works of Karl Popper, Thomas Kuhn et al., through an emphasis on the experimental practices. Thus, historians and philosophers of science, such as Peter Galison and Ian Hacking, finally gave the experimental process the place it deserved in the scientific firmament, by framing theory with experimentation and instrumentation as equivalent actors[3]. More specifically, Galison, in his seminal work *Image and Logic* (1997) argues that we should conceive of modern physics not as a theory-centric construct but as an endeavor that is composed of three autonomous yet interrelated levels: theory, experimentation and instrumentation. In fact, he goes on to claim that each level has its own unique periodization.

By expanding on Galison's train of thought, we could argue that even if one of the aforementioned levels faces a crisis, as is apparently the case with the theory level nowadays, the other levels can continue to operate with a certain degree of autonomy, in much the same way that the design of the FCC continues without being put on hold by this crisis. Obviously, the relationship between the three levels is much more harmonious during historical periods where they progress as one, as in the case of the Standard Model, where there was a specific roadmap of theoretical predictions that led to great experimental discoveries. Unfortunately, however, in practice each level's autonomous course and the relationship between them are seldom unhindered. Thus, today we find ourselves in an era where, following the discovery of the Higgs boson in 2012, the "hunt" for the SM particles has come to an end. Nevertheless, there are numerous indications[4] that the SM is not the ultimate theory of elementary particles and their interactions (Gianotti 2004, Ellis 2012). Quite a number of theories were developed in order to overcome the problems surrounding the SM, like Supersymmetry, but none were able to be experimentally verified until now. Despite all this, the scientific endeavor has shown that it can always find new, creative ways to surmount any obstacle placed in its path. And one such way is for the experimental process to acquire the leading role so that it may assist in pulling the stuck wagon of High Energy Physics (HEP) out of the mire. As one of the most respected theoretical physicists of our time, Nima Arkani-Hamed, claimed in a recent interview, "when theorists are more confused, it's the time for more, not less experiments" (Arkani-Hamed 2019).

---

[1] Grigoris Panoutsopoulos is a Ph.D. candidate at the University of Athens
[2] For more details visit the website https://fcc.web.cern.ch/Pages/default.aspx
[3] See also Arabatzis 2008

Besides, this would not be the first time that the experiment assumed a leading role in the scientific process, as quite a few experiments have historically not taken place following requirements set forth by a well-established theory, as a stereotypical view of science would assert, but were instead realized for the purposes of exploring new domains. For example, as Friedrich Steinle (1997, 2002) argued, many electrical phenomena were discovered by physicists such as Charles Dufay, André-Marie Ampère and Michael Faraday in the 18th and 19th century through experiments which had not been guided by any developed theory of electricity. According to Steinle, exploratory experiments "typically take place in those periods of scientific development in which -for whatever reasons- no well-formed theory or even no conceptual framework is available or regarded as reliable. Despite its independence from specific theories, the experimental activity may well be highly systematic and driven by typical guidelines" (1997, 70). In this context, it is necessary that experimental instruments are allowed "a great range of variations, and likewise [to] be open to a large variety of outcomes, even unexpected ones" (2002, 422). By sifting through examples of an exploratory experimentation closer to the present day, we could argue that the entire history of particle physics is full of such cases. In this way, a continuous and laborious exploratory experimental practice, which lasted for decades, has largely characterized the post-war course of HEP, and it was what allowed the SM to emerge through a "zoo" of newly discovered particles. Koray Karaca (2013), for instance, offers a very interesting case study on the deep-inelastic electron-proton scattering experiments, which were performed during the late 1960s at the Stanford Linear Accelerator Center (SLAC) and which led to the discovery of quarks as the composite parts of nucleons.

---

[4] The critique against the SM is mostly based on the un-naturalness of the model, the hierarchy problem and the issue of fine-tuning.

## The dynamics of the experimental process at CERN

Almost four decades before the recent debate surrounding the relationship between theory and experiment in the post-Higgs era, Viki Weisskopf, the former director-general of CERN and one of the most influential personalities in the field of particle physics, put forth an impressive allegory during a speech:

> *"There are three kind of physicists, namely the machine builders, the experimental physicists, and the theoretical physicists. If we compare those three classes, we find that the machine builders are the most important ones, because if they were not there, we would not get into this small-scale region of space. If we compare this with the discovery of America, the machine builders correspond to captains and ship builders who truly developed the techniques at that time. The experimentalists were those fellows on the ships who sailed to the other side of the world and then jumped upon the new islands and wrote down what they saw. The theoretical physicists are those fellows who stayed behind in Madrid and told Columbus that he was going to land in India."* (Weisskopf 1977)

We believe that one should keep at a certain distance from Weisskopf's stringent view and recognize the exceptionally important contribution of theoretical physicists for modern physics. We should also admit, however, that despite the fact he was a theoretical physicist himself, he was able to recognize the momentum that the experimental process could achieve within the context of Big Science and also how difficult it would be for theory, experimentation and instrumentation to proceed indefinitely as one. Besides, being so far removed from an experimental scale that involves experiments that can be set up on a bench, designed and realized within the span of a few months by 2-3 scientists, we should accept that the design and realization of projects such as the FCC, should have their own «timeline». Given the decades-long timeframe requirement for the realization of a collider, it becomes quite difficult to predict the possible developments in theory or in ongoing experiments in the meantime. Hence, the time scale of these projects is another factor testifying to their independent dynamics, which may not always be compatible with theoretical developments.

This is the strategy that has been followed, as a rule, by every collider in the recent history of CERN, since they are designed with a wide array of goals in mind and not just to test a very specific theoretical prediction. The most characteristic case is that of the fabled LHC. At every workshop involved in the design of the LHC (Lausanne/Geneva 1984, La Thuile/Geneva 1987, Aachen 1990), what was mainly at stake was not the confirmation of the SM but how we could move into the realm of *New Physics,* beyond the SM. As the report from the first workshop in Lausanne states: "the standard model, with all its brilliant successes, does not explain enough […] One may say that there is at present a theoretical consensus that the once fashionable desert will actually bloom, but there is no consensus on what flowers exist there" (ECFA-CERN Workshop 1984, 8, 12). Similarly, the report from the Aachen workshop (ECFA-CERN Workshop 1990, 132) claims: "Since the ultimate goal is to look for new physics, it is also necessary to include the simulation of different alternative scenarios". Even more clear regarding the openness of the design of the LHC was Chris Llewellyn Smith who, as the theoretical convener of the Lausanne workshop, during his talk argued:

> *It is less clear that it is sensible to discuss the physics that might be studied with such a machine […] A theoretical consensus is emerging that new phenomena will be discovered at or below 1 TeV. There is no consensus about the nature of these phenomena but it is interesting that many of the ideas which have been suggested can be tested in experiments at an LHC. Although many, if not all, of these ideas will doubtless have been discarded, disproved or established by the time an LHC is built, this demonstrates the potential virtues of such a machine"* (Llewellyn Smith 2015, 4).

Therefore, when quite a few present-day critics of the FCC claim that its construction is meaningless since we lack specific theoretical predictions in need of experimental validation, in contrast to the LHC and the SM, they would do well to retrace the history of the latter's design and see how it was built, more or less, as an exploratory machine. Accordingly, the FCC strives at present to retain the exploratory spirit of its predecessors. It is not intended to be used as a mere verification tool for a monolithic theoretical scenario but as a means of paving multiple alternative experimental paths for the future. In addition, we should perceive of colliders, not only as experimental setups but also as infrastructures around which entire experimental clusters are developed.

Hence, we believe that the experimental process should be allowed to develop its own momentum. Nevertheless, this does not mean that experimentation and instrumentation should abandon any effort to achieve and maintain a close relationship with the theoretical community; but one should always acknowledge and keep in mind the different dynamics and timescales involved, a prerequisite for a healthy relationship between the three. At the end of the day, there is but one physics, and it must protect its unity. It should not, however, stride towards singleness but towards a harmonious integration, allowing each of its levels the freedom to develop without suffocating restrictions imposed by the others.

At present then, when contemporary HEP are characterized more by an open-ended explorative kind of research rather than a research that has been railroaded to test any particular kind of prediction (Massimi 2019), the situation should not be regarded as unprecedented. The fact that this particular practice is not a genuine *terra incognita* does not of course mean that there exist ready-made patterns for us to follow. The path towards *New Physics* will be long and arduous, something which becomes apparent when looking at the numerous unsuccessful attempts through the years. It is, though, a path that we need to soldier through, resisting the temptation to accept simplistic models concerning the relationship between theory and experiment, as convenient as they may be. The maxim that theory always has the lead and the experiment necessarily follows, is nothing more than an erroneous conception that attempts to rein in the unending creativity of the scientific endeavor and to confine it to predetermined movements. As

P. Galison (1987, 269) puts it, we have to "step down from the aristocratic view of physics that treats the discipline as if all interesting questions are structured by high theory."